# Fusion Divided: What Prevented European Collaboration on Controlled Thermonuclear Fusion in 1958

***Abstract:*** *The European Organization for Nuclear Research (CERN) in Geneva is renowned for operating the world's largest particle accelerator and is often regarded as a model of high-profile international collaboration. Less well known, however, is a key episode from the late 1950s, when CERN was confronted with the research priorities of similar organisations. The issue centred on a CERN-sponsored study group on controlled thermonuclear fusion, which brought together scientists from CERN member states, as well as representatives from the European Atomic Energy Community (EURATOM), the European Nuclear Energy Agency (ENEA), and the US Atomic Energy Commission (AEC). While the CERN Study Group on Fusion Problems succeeded in creating an international network for exchanging reports and coordinating projects to avoid duplication, it ultimately failed to establish joint fusion research programmes. This article explores the reasons behind this outcome to provide insights into intergovernmental power dynamics, underlying competition, and how these factors favoured the creation of a new fusion research institution in the UK, the Culham Laboratory. In doing so, the article contributes to a deeper understanding of the role of science in European integration, while also highlighting that CERN's involvement in application-oriented research remains an underexplored aspect of its history.*

Building a fusion reactor represents a 'super problem', argued the Norwegian physicist Svein Rosseland in 1958—one requiring worldwide collaboration between governments and research institutions and, in the final phase, the 'construction of pilot plants'.[1] The International Thermonuclear Experimental Reactor (ITER), under construction in southern France since 2007 to explore fusion reactions as a source of energy production, seems to epitomise the kind of large-scale project Rosseland had in mind. However, ITER is merely the latest prominent example of how fusion science is driven by the hope for an energy transition in the future and thus the quest for optimal technology in the present, all within the framework of international scientific collaboration.[2] This article examines a precursor to ITER, focusing on a scheme initiated by the European Organization for Nuclear Research (CERN), of which Rosseland was a Council member when he articulated his vision. It shows that in the late 1950s, the idea that increased cross-border collaboration between European countries would accelerate the development of controlled thermonuclear fusion faced significant obstacles. Ultimately, plans for closer cooperation were abandoned, as fusion partners divided over the best way forward. Thus, the article offers an opportunity to examine an instance of failure of a major science diplomacy initiative.[3]

CERN was an early contributor to discussions about establishing a research facility dedicated to the development of fusion energy. Founded in 1954 as Europe's centre for high-energy physics, it now houses the world's largest accelerator complex, focused on studying particles, forces, and matter. Its convention restricts all work to fundamental research with no military applications.[4] Its fusion project is a lesser-known chapter in its history that warrants greater attention, particularly as the organisation has since become a prominent model of political integration and international scientific exchange.[5] This article recalls that in 1958, the CERN Directorate proposed a collaboration with the newly established European Atomic Energy Community (EURATOM) to assess how best to advance fusion science. Shortly thereafter, representatives from the OEEC's newly formed European Nuclear Energy Agency (ENEA) joined the inter-organisational effort. Soon after, the US Atomic Energy Commission (AEC) sent delegates to contribute to the discussions, which aimed to develop recommendations for the future of fusion science in Europe. Historian Dominique Pestre has examined this initiative within the broader context of CERN's institutional history, noting that, after several discussions, the topic was set aside in favour of a decision to focus efforts on developing the next generation of accelerators, in alignment with the original scientific mandate.[6]

Why did CERN even get involved in exploring the potential of fusion? The epistemological backdrop was a crisis in the emerging field. In the early 1950s, fusion research was conducted in secret in Britain, the United States, and the Soviet Union, primarily to develop hydrogen weapons.[7] In preparation for the Second Atoms for Peace Conference in 1958, the governments of these three countries permitted the publication of research on controlled fusion processes and related technologies, with the aim of developing applications beyond the military domain.[8] However, a key requirement for achieving controlled fusion power is the creation of stable plasma (hot, ionized gas). Scientists had understood that temperatures exceeding one hundred million degrees Celsius were necessary to sustain the process, but their attempts failed.[9] As historian Joan Lisa Bromberg has summarised, the discussions and publications at the 1958 Atoms for Peace Conference made scientists aware that their expectations for a breakthrough had been overly optimistic. And because fusion research had previously been classified, much of the work had been duplicated. Consequently, the concepts presented at the conference were strikingly similar, all relying on strong magnetic fields to hold the plasma in place and achieve the necessary density and temperature conditions (so-called magnetic confinement).[10] Realising they were far from achieving their goal, fusion scientists began revising the theoretical foundations. This shift explains CERN's involvement.

But what was the outcome? This article sheds new light on the importance of CERN's initiative in fostering exchange networks that were distinct from those associated with the European centre. While it demonstrates that CERN's Study Group on Fusion Problems was instrumental in creating loose international contacts, it also reveals that, for many years, there was neither close collaboration nor the establishment of a joint fusion laboratory in Europe. Studies of science diplomacy often portray science as a problem-solver that benefits from collaboration, thereby overlooking the rivalry, competition, lobbying, and inequality inherent in international exchange. As a contribution to the growing body of critical historical literature, this article examines the pursuit of national advantage *within* international organisations to offer insights into intergovernmental power dynamics.[11]

Drawing on archival sources that document both official positions and backstage lobbying,[12] the article explores how CERN, EURATOM and ENEA joined on fusion and how their efforts were undermined by national interests, offering conclusions on the competition for leadership. Notably, these dynamics benefited a new fusion research facility in the UK, the Culham Laboratory (now Culham Centre for Fusion Energy), founded by the Atomic Energy Authority (UKAEA) and built on a former airfield near Oxford between 1960 and 1964. In other words, CERN's brief involvement in fusion science can be seen as the starting point for a larger initiative, especially when considering the activities of the British accelerator expert John Bertram Adams. After working at the Atomic Research Establishment at Harwell, Adams joined CERN to become head of the Proton Synchrotron Division in 1954. He was entrusted with the construction of Europe's first big accelerator, which began operations in 1960.[13] Adams was not only associated with CERN, but also an influential figure in the UK nuclear programmes. Under his leadership, fusion knowledge exchange began to flourish, and this continued when he became the first director of the Culham Laboratory—a move that symbolised the rivalry between national and international projects.

In three sections, the article emphasises how past futuristic goals failed to materialise. Collaboration was hampered by the fact that some members of CERN's management sought inter-organisational exchange, while others were determined to avoid it. Ultimately, fusion science was not to become a focus of CERN, yet it was not entirely abandoned. This prompts questions about CERN's stance on applied research.

### *The dilemma of collaborating with other international organisations*

CERN's Study Group on Fusion Problems facilitated the exchange of ideas among physicists at a time when it had become clear that the anticipated progress in fusion energy for civilian applications had been overestimated. The aim of the group was to provide an overview and make proposals for future action six months after the Second Atoms for Peace Conference, held in September 1958 at the UN Palais des Nations in Geneva, Switzerland. Prior to the conference, the UK, Soviet, and US governments announced the declassification of their fundamental research findings.[14] Their secret programmes on weapons technologies had been accompanied by research into civilian applications. As a result, all fusion science had remained classified, which hindered the exchange of findings and created asymmetries between the 'haves and have-nots' of information and infrastructure.

Certainly, the strong ties between fusion science and military interests—including the British cabinet's 1954 decision to develop its own hydrogen weapons arsenal and a first bomb test in 1957—continued to impede the expansion of international collaboration, despite the Atoms for Peace rhetoric.[15] Notably in 1958, only research involving the four primary devices of magnetic confinement (the pinch, tokamak, stellarator, and magnetic mirror) was declassified.[16] While exchange in this area expanded in subsequent decades, research in the second major branch—inertial confinement, using particle-beam energy or pulses of laser—remained classified due to its close connection to the physics of thermonuclear weapons.[17]

The partial lifting of security restrictions, and the open discussion on the state of fusion science, was a watershed in its history, sparking significant media attention. The announcement of the 1958 Atoms for Peace Conference displayed the potential of fusion energy. New knowledge on fusion processes was propagated as serving the future of humanity, with fusion reactors being presented as the solution to the growing energy demands. The three leading fusion nations showcased their progress in plasma production, with scientists presenting talks based on empirical data and theoretical reflections.[18] Yet the actual conference content proved embarrassing, as many of the predictions were shown to be false. Upon his return, a member of the Harwell team (the UK's chief civilian nuclear research hub), reported that his 'outstanding impression of Geneva' was 'that we might as well dismiss as fantasy any idea that there is a short cut to a fusion reactor', while a colleague expressed 'mild pessimism' and another called for an increase in theoretical efforts.[19] A major disappointment was their own device, the toroidal pinch ZETA (Zero Energy Thermonuclear Assembly): the plasma was confined far less effectively than originally thought, and instability remained (see figure 1).

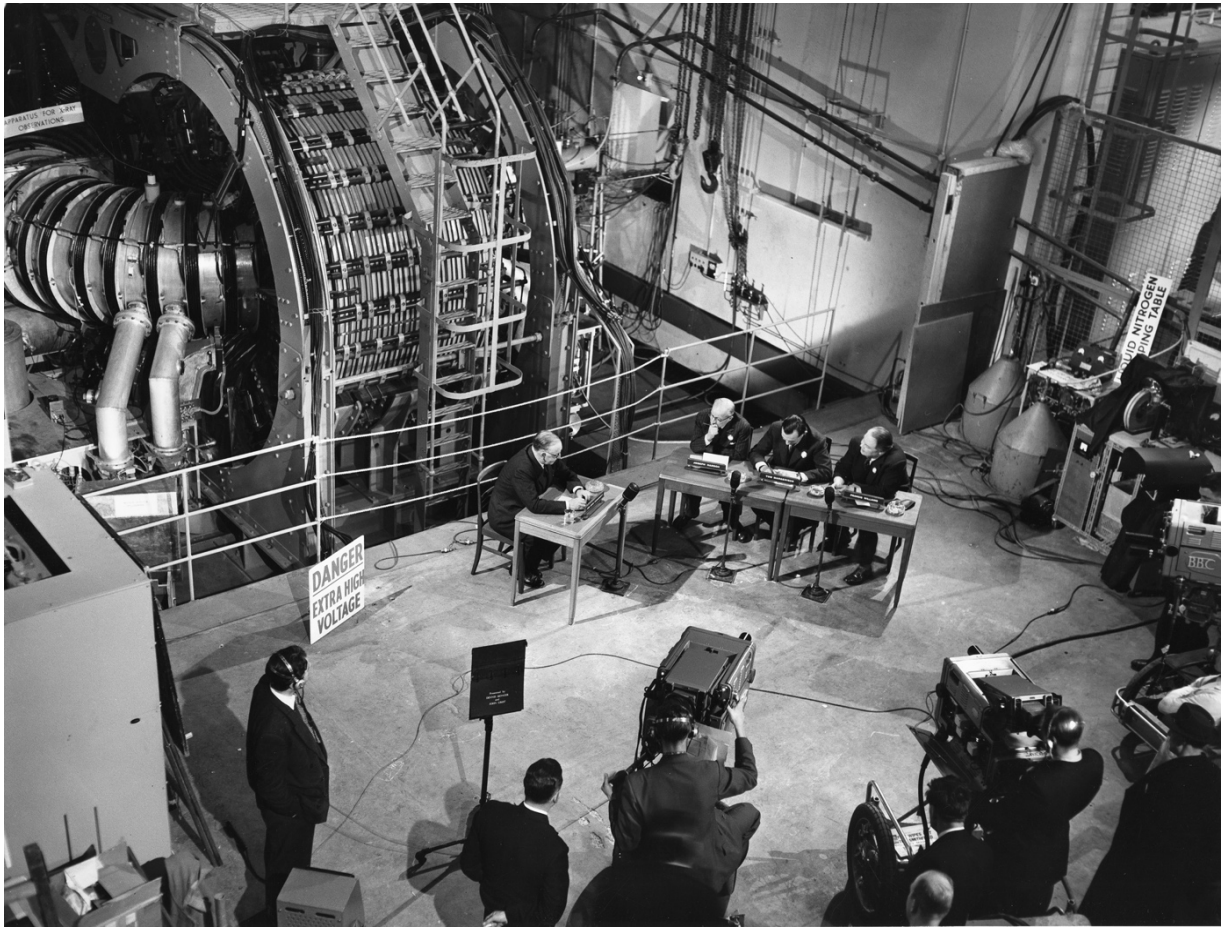

Figure 1. John Cockcroft (1897-1967), Director of the Atomic Energy Research Establishment at Harwell, interviewed by a team of reporters about ZETA, shown at the upper left. He provided an optimistic assessment of fusion energy but was soon compelled to publish a retraction (created by the UK Government in the public domain, copyright expired)

Given its size and funding, CERN was the organisation best positioned to host new plasma physics programmes and an experimental facility for such fundamental studies. By 1958, it already had a team of physicists studying plasma, as evidenced by the request to deploy a British scientist to work on this subject and the progress report submitted by the Proton Synchrotron Division.[20] CERN's initiative to explore fusion potential developed through the collaboration of its Council (the governing body composed of country representatives), its Scientific Policy Committee (the advisory board of leading physicists), Cornelis Bakker, a cyclotron expert who had been Director-General of CERN since 1955, and John Bertram Adams, head of the Proton Synchrotron Division and the main driving force behind the effort.

Adams, who had gained the necessary expertise in using magnetic fields and electrical engineering techniques for fusion through his work on accelerators, held a meeting with several European researchers in March 1958, before the Second Atoms for Peace Conference. The meeting participants discussed results from Adams' former colleagues at Harwell, particularly those working on the ZETA.[21] They summarised experiments and shared information on laboratories either already involved in fusion research or considering entering the field. Adams concluded the meeting with the proposal to arrange a series of follow-up discussions.[22]

In a letter to Bakker, Adams proposed the names of fourteen European physicists he wished to involve in a more formal study group. Among them were himself, Arnold Schoch, and Jiri George Linhart, who had joined his division at CERN in 1953 and 1957, respectively, to develop accelerator concepts.[23] Some of the interests of accelerator experts and fusion scientists overlapped: Schoch and Linhart were exploring new possibilities at the intersection of high-energy and plasma physics. Their interest stemmed not only from the particles discovered at the ZETA, but also from the realisation that a thorough examination of the plasma-physical processes in accelerators could contribute to advances in their design.[24]

These scientific developments were accompanied by diplomatic negotiations to garner support. Adams' idea to hold further meetings developed in parallel with an initiative by the French diplomat François de Rose, President of the CERN Council, who saw the laboratory as a potential candidate to conduct research under contract with EURATOM. This organisation, formed in 1957 by the six members of the European Coal and Steel Community, had the specific mandate to foster nuclear power development for energy production purposes.[25] De Rose had been approached by Louis Armand, President of EURATOM, who, three years earlier, had provided scientific advice to the Organisation for European Economic Cooperation (OEEC, now OECD).[26] In his new position, armed with a significant budget, Armand explored options within the broader nuclear research field and expressed interest in forming links with CERN, although he did not specify the exact area. On the advice of Bakker, de Rose suggested plasma physics. Bakker saw its inclusion in CERN's programmes as a way to retain engineers after the completion of the Proton Synchrotron, thus broadening CERN's focus beyond high-energy physics.[27] His concern aligned with Adams' worries about extending the contracts of talented senior scientists in the Proton Synchrotron Division once this accelerator had been completed. Two influential members of CERN's management, the British physicist John Cockcroft and his Italian colleague Edoardo Amaldi, supported these views, alarmed by the growing number of staff leaving the laboratory in Geneva to pursue academic careers elsewhere.[28]

The loss of skilled personnel could weaken CERN at a time when it was striving to catch up with the US National Laboratories in terms of technical achievements and scientific excellence. In 1958, the Proton Synchrotron was still under construction, but it was clear that it would be operational in two years. With its completion (and the earlier Synchrocyclotron), the original CERN mission was fulfilled, even though discussions were already underway about a third machine.[29] As the next organisational goals had not yet been defined, the laboratory entered a phase of experimentation. CERN's exploration of the plasma physics option must be understood in the context of an uncertain future, when, simultaneously, the European centre had to contend with the interests of other institutions regarding research priorities.

The various strands of discussion regarding CERN's future, underpinned by concerns about the future of work, culminated in the proposal for a 'EURATOM-CERN Joint Study Group for Fusion Research'. In May 1958, Adams and Bakker met with Jules Guéron, EURATOM's General Director of Research and Education, who had previously headed the French nuclear research centre at Saclay. Together, they developed the idea of creating an overview to be submitted to both organisations by the end of the year.[30] Guéron was authorised to take all necessary steps. At CERN the proposal required the support of its management, composed of representatives from all member states. In May 1958, its Scientific Policy Committee reviewed the proposed inter-organisational collaboration. Notably, the joint study group was now also tasked with recommending future programmes that could be conducted 'either by existing national research centres or by further development of some of these centres or by the creation of a European centre'.[31]

However, it was not this goal that raised reservations among CERN's scientific advisors, but rather the question of whether CERN should be involved in principle. The German representative, Werner Heisenberg, pointed out that plasma physics would require a new, costly machine, while the UK representative, Patrick Blackett, argued that it went beyond CERN's basic research remit, potentially leading to patent problems. A fusion study programme would require careful consideration to ensure it did not conflict with CERN's convention. In response, Bakker stated that 'any co-operation […] should not affect the research work with the accelerators and should not change CERN's policy of being a completely open institute'.[32] After the committee's chairman, Edoardo Amaldi—who also served as the first chairman of EURATOM's Scientific and Technical Committee—argued that practical applications could be excluded from the study, the advisory board agreed on the formation of a joint study group that would include EURATOM staff, as well as scientists from CERN's member countries and the centre itself.

Although the CERN Council had to make the final decision, Bakker now optimistically prepared a press release highlighting that CERN, in collaboration with EURATOM, was about to establish a study group addressing the fundamental problems of plasma physics related to fusion. However, instead of informing the press, following the Council meeting on 20 June, Bakker had to tell Guéron that its members would reconvene for an extraordinary session a week later.[33] This was because the representatives of the non-EURATOM states had vetoed the proposal, so it had to be discussed twice (with non-members at CERN including Switzerland, Greece, Yugoslavia, the UK, Sweden, Norway, and Denmark; and members the Federal Republic of Germany, France, Italy, the Netherlands, and Belgium).

National imperatives now started to impact on the fusion collaborative venture. The Italians and French, aware that their countries were lagging behind the three leading fusion nations, including the CERN member Britain, sought to take advantage of the 1958 relaxation of secrecy to foster collaboration among Europeans. In contrast, the British Donald Fry criticised the extent of the collaboration and called for it to be limited to an informal survey.[34] Fry, a radar pioneer and the first director of the Winfrith Atomic Energy Establishment—constructed from 1957 to host several experimental fission reactors—was likely motivated by concerns over any potential expansion of fusion programmes. However, behind his objections lay broader British motives. International tensions had already arisen regarding the necessity of EURATOM when first plans were discussed in 1955. In this context, the UK sought to protect its interests by limiting information-sharing to prevent others from becoming competitive, especially in uranium enrichment and weapons design. The UK sought exchange at the intergovernmental level and did thus not support integration by opposing EURATOM while backing the more modest project proposed by the OEEC for establishing the ENEA.[35]

The suggestion of a CERN fusion project involving EURATOM put the British in an awkward position. Representatives of other non-member states also resisted the plan. At the conclusion of the Council discussion, the Swiss, the Greeks and the Scandinavians proposed an amendment affirming that non-EURATOM states would not commit. Since the invitation of the OEEC could help circumvent the issue of non-membership, the Swiss proposed discussing the results within this framework.[36] In fact, Pierre Huet, Director-General of the OEEC's newly established Nuclear Energy Agency (ENEA), inquired about the possibility of joining the inter-organisational effort and suggested that non-EURATOM states could participate through the ENEA.[37] Given the wide range of opinions and options, CERN Council President de Rose postponed the discussion until 27 June.

Between the two meetings, the UK delegation continued to oppose the plans. While recognising the need to assess the status of fusion research following the declassification of numerous reports under *Atoms for Peace*, they supported the idea of a study group—if it excluded EURATOM. They also questioned whether fusion even fell within CERN's objectives. Fry informed Blackett that such a study might place an undue burden on their senior staff.[38] H.L. Verry, who advised CERN on financing, wrote to Bakker, noting that the Council was free to decide how and why it would collaborate with any organisation.[39] However, letters to his compatriots revealed that Verry's view of collaboration with EURATOM was, in fact, negative. Already after the March meetings, Verry had argued that a joint effort could set an 'embarrassing precedent'.[40] In subsequent discussions, the UK opposition would gradually bring the CERN fusion initiative to an end. Cockcroft, who had advised the UK government in 1951 on the need for a concentrated national fusion science programme, was quick to recognise that 'it is clear that the United Kingdom cannot stand aside and leave the decision to other people in the Council'.[41] The inclusion of plasma physics and fusion science in CERN's research agenda was not something he supported. As early as 1957, when rumours circulated about expanding CERN's work, Cockcroft argued that it would alter the laboratory's character.[42] Nevertheless, he acknowledged that the United States 'would welcome the entry of CERN into the field of plasma physics' since it was 'a supporter (honorary) both of Euratom and CERN'.[43] This stance, at least, opened the possibility of extending the desired exchange across the North Atlantic, particularly given the US lead in the nuclear field.

The Norwegian and Danish delegates aligned with the British position, which strengthened the influence of the non-EURATOM faction at CERN.[44] Cockcroft reassured the Norwegian representative that he was uneasy about working with EURATOM and did not support a joint project. The Danish delegation suggested that an overview be conducted in various countries and then presented to the CERN Council.[45] In their additional resolution, the British made it clear that any potential fusion study group would only review the current state of affairs and offer recommendations. The group would neither conduct research nor issue binding recommendations. Nevertheless, they amended the proposal to allow both EURATOM and ENEA to send observers. On 27 June, the CERN Council adopted these proposals.[46]

While officially seeing the UK delegation's role as that of a mediator, Verry privately remarked that their efforts could be perceived as a 'sabotage of the French and Italians interest' in collaboration. He argued that the British had only agreed to consider alternatives to high-energy physics because 'in the absence of fusion research or some other attractive toy, some of the CERN staff will be lured away as they see the accelerator programme being completed'.[47] The British, a leading nation in fusion science, where information was only just being released for the exploration of civilian uses, supported the creation of a targeted study group under CERN's auspices due to unresolved questions about the future direction of work. They agreed, although they had a strong interest in ensuring that CERN remained focused on high-energy physics.[48] The specific assignment of tasks can be seen as a proposal to complement institutions working on similar subjects. However, it is more plausible to conclude that it was intended to exclude any competition to national programmes.

*Plasma physics at CERN? Or a second European laboratory?*

In July 1958, Bakker and Adams informed Guéron that the CERN Council had decided not to work with EURATOM on a joint study. While this decision involved minor financial sacrifices, it had little practical impact, as delegates could still attend the meetings and receive all reports.[49] The decision limited EURATOM's influence, favouring CERN's position rather than offering equal footing. Under the chairmanship of Adams, CERN began conducting a survey that had not been predefined in terms of a purpose—whether it was to provide a scientific overview for other organisations to take action, develop proposals for CERN to expand its own programmes, or define how to establish a second European laboratory. Despite the lack of clear objectives, the initiative bore much resemblance to the arguments made when CERN was founded just a few years earlier, notably that European science lagged behind others and that considerable financial investment was required to catch up. The activities were further fuelled by the realisation that, after initial hype, the construction of a functional fusion reactor for successful energy production remained a distant prospect. Consequently, participants sought to gain an overview, summarise existing results, and evaluate methods for achieving fusion to develop new scientific ideas and make policy recommendations.

The CERN Study Group on Fusion Problems met three times in Geneva until March 1959, with the Second Atoms for Peace Conference serving as the starting point for discussions on the content and direction of fusion science in Europe. Invitations were sent directly to recognised experts, rather than to country representatives. Adams took a strong lead, introducing and guiding the discussions, as well as circulating the meeting minutes. It is therefore fair to conclude that he built and expanded his own scientific network, with attendees eager to engage in discussions 'with the top-grade scientists working in the plasma physics field throughout Europe'. In fact, the group consisted solely of CERN member states with the capacity to support the initiative.[50] Their science administrations received individual reports from their delegates.[51]

In addition to participants from CERN and its member states, Guéron attended the first meeting, joined by Donato Palumbo, who had been entrusted with EURATOM's new fusion programme. Pierre Huet, the Director-General of ENEA, delegated his science advisor Lew Kowarski—who was also a senior staff member at CERN.[52]

The first meeting, held in September 1958, aimed to assimilate information from the Atoms for Peace Conference, focusing on the technical and methodological approaches to magnetic confinement and plasma stability. As insufficient time had passed since the general release of information, participants agreed to undertake surveys on devices and diagnostic techniques. A notable suggestion was to read the book by US fusion specialist Amasa S. Bishop, *Project*

*Sherwood*.[53] Bishop, who had previously led the Research Division of this classified programme to develop controlled thermonuclear fusion, and in 1956 became the technical representative of the US Atomic Energy Commission (AEC) in Europe, working at the embassy in Paris, was invited to the second group meeting in December. The absence of open discussion at the CERN Council regarding both the role of the AEC and Bishop highlights the significant influence of American fusion science. Adams indeed concluded that Europe would need to support research on a scale similar to that of the United States to make a meaningful contribution. As part of the study, comparative data on financial investments in fusion science from the 'free world outside USA' was thus collected to inform recommendations.[54]

The second meeting, attended by 36 delegates, aimed to evaluate surveys, discuss existing programmes, and assess the possibility of a joint centre. Interestingly, Adams recalled that the ultimate goal was the production of fusion energy, and that governments were less willing to support only plasma physics,[55] which indicates an approach that ran counter to CERN's core objectives. Participants carefully evaluated competing methods and presented the status of their research. From this, Adams drafted a summary report submitted to the CERN Council for its May 1959 session. The draft was discussed at the third meeting in March. In it, Adams detailed that most work in Europe was still in the planning stages, making it essential to consider how programmes could be interlinked. The question of a joint laboratory remained the most controversial issue. While CERN was viewed as a model of European integration—an idea that influenced deliberations, particularly since the United States had not centralised fusion science in a single laboratory—the recommendation differed from that made for high-energy physics several years earlier. The study did not propose the establishment of a new laboratory, nor did it deem supranational efforts necessary, nor recommend a concentration of efforts.[56]

However, this conclusion did not reflect all opinions. The Norway delegate Kjell Johnsen, who had worked in Adams's division at CERN but had been a professor in Trondheim since 1957, argued to Adams that the Scandinavians feared CERN's fusion programme would undermine high-energy physics. Johnsen considered that 'European fusion research on a big scale should be independent of CERN'. However, he added that he would not object 'if CERN's scientific and administrative experience can be drawn upon in building up a European laboratory on fusion', also if the CERN plasma physicists were made the core group of this centre.[57] But neither of these ideas came to fruition.

One explanation for the failure to initiate a European fusion centre is that CERN ultimately withdrew its support. Before the Council session in May 1959, members of the leading board, now including Director-General Bakker (see figure 2), suggested that CERN should not be involved in a field with commercial and strategic significance, echoing concerns raised by the Scientific Policy Committee the previous year. The board recommended that another European organisation expand its aims to cover fusion science.[58] CERN narrowed its focus back to its key mission, engaging the engineering staff in the development of a bubble chamber and the accelerator designers in improving the existing complex. This resolved the issue of continuous employment, while the need for a larger accelerator in the future became increasingly evident.[59]

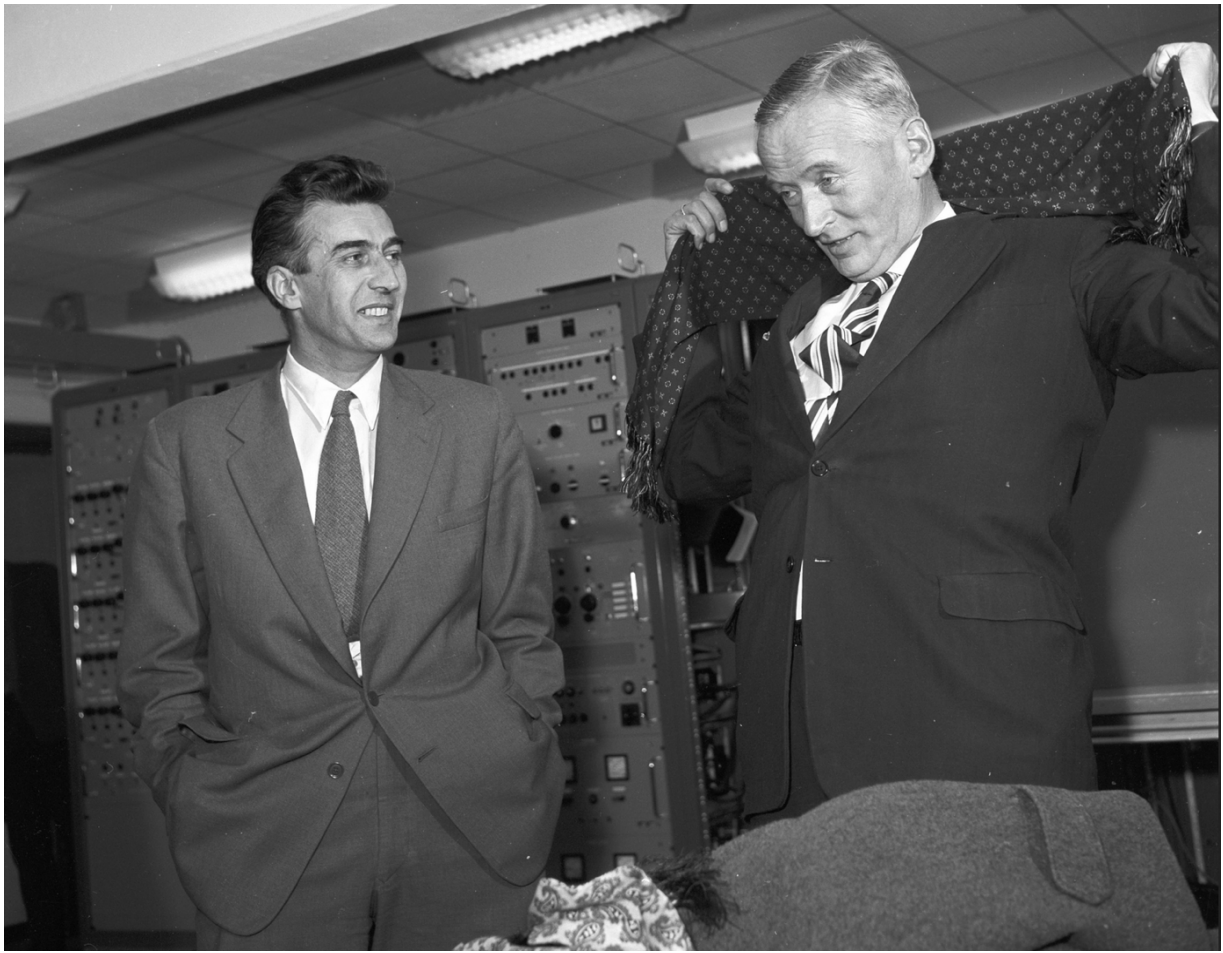

Figure 2. John Bertram Adams (1920-1984) and Cornelis Bakker (1904-1960) in the control room of the Proton Synchrotron during an interview on 15 January 1960 (Courtesy © CERN, Reference: 2432)

The question was not about how to support fusion science, but whether to support it at all: The emerging field competed for limited human resources and money, meaning that dividing them between two European units could reduce what was available for CERN. While the Council did not advocate for a new laboratory, members still supported the French representative Perrin's view that it was worthwhile to consider a European-level solution.[60] Given CERN's leadership in evaluating a potential international fusion effort, the Council agreed to sponsor the continuation of the study group until an alternative solution was found. The main option discussed was the establishment of a society called the European Study Group on Fusion, with the same objectives: offering meetings and exchanging reports, allowing participants 'to plan their research programmes with full knowledge of the activities of other laboratories.'[61] Organisational and scientific reasons explain why CERN became involved in assessing the state of fusion and proposing future policy, but interest in the initiative waned dramatically. At its December 1959 session, the Council concluded that CERN was no longer making a significant contribution. To do so, it would need 'to maintain a staff actively engaged in fusion work, which was likely not to be the case, as the plasma work at present being done in the accelerator research group will gradually tail off'.[62] The research led by Linhart was expected to end soon, and the Council was unwilling to consider a new programme and provide support.[63] Plasma physics had been a part of CERN's activities but was discontinued when, in 1960, Linhart became the director of a EURATOM fusion research group at the Frascati Laboratory in Rome, Italy.[64] At this stage, national interests prevailed over collaborative plans, and those eager to continue the exchange were forced to reconvene under EURATOM. However, Frascati never became for plasma physics what CERN was for high-energy physics.

*An informal European network and a new British laboratory*

In a second report to the CERN Council in 1962, Adams underscored the difference to high-energy physics, arguing that no case could be made for 'a European organisation for fusion research that would be sufficiently strong to carry such a proposal through the long stages of governmental procedure'.[65] The bitter pill of dividing efforts in the pursuit of fusion power was somewhat sweetened by scientific considerations: from the outset, Adams argued that instead of embarking on large-scale reactors, the focus should be on attacking plasma physics to later identify promising devices. Given that the basic principles of thermonuclear reactions were not fully understood, scientists decided to explore plasma behaviour further. They prioritised exploring competing methods and techniques, with a fusion reactor for energy production remaining a distant, common goal.

Adams also pointed out that fusion was inextricably linked with nuclear energy and those organisations established to develop and exploit nuclear energy from the fission process. ENEA-led support now concentrated on fission, as evidenced by the commissioning of two experimental uranium reactors: Halden in Norway (1958) and Dragon in Britain (1959).[66] Kowarski, ENEA representative and CERN staff member, argued against Europe competing with nations most actively engaged, advocating for complementary efforts.[67] In the fusion field, the drive for national advantage ran counter to the ideals of a central unit. As historian Barbara Curli has noted, from 1959 onwards, EURATOM awarded association contracts to national laboratories.[68] However, coordination and central financing marked only a small step forward. There was no European integration in this area and era. The preferred approach to fusion science was one of competition and complementation, rather than collaboration.

As enthusiasm for international collaboration waned, Adams' network continued to serve as a key point of contact. In 1959, he chaired two more meetings, which were no longer held at CERN but at national institutions that had established, or were planning to set up, fusion research programmes. At the fourth meeting in Harwell, the 45 participants were informed that the CERN Council no longer supported their meetings but would fund the necessary secretary work. A month before the next meeting in November 1959 in Munich, where the Max Planck Institute for Plasma Physics was soon to be established, Adams informed Cockcroft that the group would continue informally.[69]

In parallel with sustaining the informal exchange forum, national plans were developed further. In other words, another reason the CERN initiative failed to provide a solid foundation for deeper collaboration was the growing emphasis on national control. Particularly the British opposition proved insurmountable. The UK gradually reduced its support for the original CERN initiative in the charged context of re-rapprochement with the United States in the military domain. Only a week after the CERN Council had decided to establish its Study Group on Fusion Problems, the US and UK governments signed their Mutual Defence Agreement (MDA) on 3 July 1958. This restored their nuclear collaboration while causing friction with the European allies, who had only recently formed EURATOM. Fusion power served both military and civilian needs, leading to the delicate balance between secrecy and the sharing of knowledge.[70] In the military domain, the UK sought alignment with the US while in the civilian domain, it sought influence in continental Europe without offering too close an exchange. Adams' network facilitated contact between British researchers and their colleagues abroad.

Meanwhile, recognising the need for more basic research with new equipment, the British merged some of the existing fusion programmes at Harwell (relieving it from expansion) and Aldermaston (founded in 1950 for A-bomb research and expanded in 1954 to include work on H-bombs) into a single scheme with non-classified work. The new laboratory enabled international collaboration and facilitated the hiring of scientists from abroad. Its aim was to develop controlled fusion of light elements to establish whether a fusion reactor was a practical possibility.[71] In July 1958, alongside the signing of the MDA with the United States, it was proposed to transfer the civilian part of the fusion programme to Winfrith under the direction of Fry. However, the UKAEA revised these plans 'in view of Geneva developments' after the disappointing outcome of the Second Atoms for Peace Conference. In December, it was recommended that a separate facility be set up, soon named Culham Laboratory.[72]

A month earlier, Adams, the central figure in assessing Europe's fusion potential, had begun serving on the UKAEA's controlled thermonuclear fusion research advisory committee, a body chaired by Cockcroft.[73] Cockcroft had supported Adams' career since his time at Harwell, and now Adams was one of the employees CERN was about to lose. In August 1959, plans for Culham came under his wings as director-designate, with his return to the UK expected in October 1960.[74] However, following the accidental death of Bakker in April, Adams was nominated interim Director-General of CERN. As a result, he divided his time between the two laboratories until August 1961, when the Austrian-born US physicist Victor Weisskopf was appointed Director-General of CERN.[75]

At the Council meeting in December 1960, President de Rose emphasised that the French were now satisfied with the useful connection that the (now renamed) European Study Group on Fusion offered. Given the decision that Adams would leave, Edoardo Amaldi asked who would now be in charge of this CERN-sponsored endeavour.[76] Although the original goal of gathering information for the Council had been achieved, members decided to continue supporting the group for another year and approved a special agreement between CERN and Adams to maintain contact. The same occurred a year later, in December 1961, when the new director, Weisskopf, preferred to leave the sponsorship with an international organisation, even though he argued that CERN's interest in fusion science was limited.[77]

CERN never integrated fusion science into its policies or programmes, nor did it develop a fully-fledged plasma physics project. The activities facilitated by the informal network gradually became obsolete as similar arrangements were established. In addition to EURATOM, the International Atomic Energy Agency (IAEA) in Vienna began hosting conferences, the first in 1961.[78] CERN had pioneered this format in Europe, but the last time any activity was mentioned at its Council meetings was in June 1962, when Adams' second report was acknowledged.[79] Archival records suggest that the group was disbanded in 1964 after a total of ten meetings, the last five held annually: at Fontenay-aux-Roses in Paris (1960), the Frascati Laboratory in Rome (1961), in Amsterdam (1962), Stockholm (1963), and at the Nuclear Research Centre in Julich (1964).[80] The group's dissolution likely coincided with Adams' involvement in UK politics after his appointment as Controller at the Ministry of Technology in July 1965, soon to become Member for Research of the UKAEA. A year earlier, his activities had attracted the attention of an editor working on a book on international organisations, but Adams declined the offer to include the European Study Group on Fusion, remarking that it was 'more or less a private club with no offices, no officers and no funds'.[81]

*Epilogue: CERN's hidden history of fusion*

CERN's involvement in the controversy surrounding the best way to support fusion science provides insights into how international collaborations can fail to materialise when competition outweighs the desire to establish exchange. In 1958, national interests impeded the possibility of a CERN-like fusion centre. The first attempt at European integration lasted only a few years and primarily served as a forum for physicists to share progress reports. The meetings saw a high turnover of participants (total 250), with a core group of ten regular attendees. Ultimately, the group never developed a unified strategy. Despite CERN's potential to serve as a coordinating body or host institution, its Council resisted supporting plasma physics and opposed the creation of a second European research centre. CERN's management had endorsed the study group for utilitarian reasons: The centre was grappling with the direction of its scientific mission as its planned accelerators neared completion. Its management agreed to explore the possibility of expanding into new scientific fields, but soon abandoned this idea in favour of maintaining the core focus on high-energy physics. This decision must be understood in the context of competition for funding and competition within physics over which subjects were deemed relevant—of which the fusion controversy is just a case in point. Underlying the scientific debate, however, was especially the opposition from UK representatives at CERN to expanding European integration to a degree that would threaten their country's privileged position as the US's nuclear defence partner. This resistance stemmed from concerns over having to share too much fusion expertise, both in the context of the arms race and the pursuit of economically advantageous civilian applications.

The history of unsuccessful attempts at scientific collaboration, trial-and-error exchange, and the way national interests influence the activities of international organisations might suggest a narrative of failure, but this is far from the conclusion this study aims to draw. In fact, the CERN initiative planted important seeds. One key outcome was the decision to test and develop existing concepts and devices of magnetic confinement competitively, with the goal of identifying the most effective solution. Secondly, the group provided an initial forum for Europeans to discuss fusion science at a time when it had not yet emerged as a distinct subdiscipline and when there was no specialized training. This underscores the significance of the ten meetings, as they enabled professional contacts that, alongside the scientific exchange, may have been crucial for the careers of those involved, thus having long-term effects.

There was a growing demand for expertise as the number of fusion research facilities increased. From 1959 onwards, in the UK, the Culham Laboratory focused on controlled thermonuclear fusion, while research in continental Europe took place at new institutes: among them the Frascati Laboratory in Italy, the Max Planck Institute for Plasma Physics and the Julich Nuclear Research Centre in Germany, the CEA Laboratory in Fontenay-aux-Roses in France, the Risø Research Centre in Denmark, or the EPFL Plasma Physics Laboratory in Switzerland. Association agreements with EURATOM gave access to research funding and facilitated collaboration, a policy that was later extended to include non-member states.

As stressed, EURATOM, which had initially sought a joint effort with CERN to explore fusion options, took on a leadership role in coordinating research efforts across continental Europe. This arrangement remained in place for over a decade. In 1968, Adams revisited the idea of a joint enterprise. In a letter to his successor at Culham, Rendel Sebastian ('Bas') Pease, who had been one of the participants in the original study group, Adams noted that ten years earlier, a proposal for a single laboratory had failed to gain the support of enough countries. Pease responded that while a solid supranational collaboration would strengthen fusion science, there was 'no pressing need for a single central facility such as that at CERN'.[82] Two years later, however, the UKAEA reviewed its international exchange and concluded that closer collaboration was becoming important, 'as the fusion programmes reaches the point at which really large-scale experimental equipment are required'.[83] In a 1973 lecture, Pease noted that the United States and the Soviet Union were conducting research on a broad basis, while smaller nations had to work on a cooperative basis with disproportionate investment.[84] Such conclusions reflected a broader trend towards centralisation, after the previous decade's emphasis on competitive scientific approaches had proven unsatisfactory. That same year, 1973, the framework for British participation in EURATOM was formalised with the UK's entry into the European Community. It marked an important shift in policy in the UK. The creation of the Joint European Torus (JET) in Culham then enabled the first collaboration on fusion hardware and design. For more than three decades, JET shifted the principle of exchange towards European integration, including the UK.[85]

As noted, CERN discontinued its limited support of plasma physics in 1960. However, in his second report to the CERN Council in 1962, Adams criticised the lack of exchange between accelerator and fusion researchers, despite both fields sharing concerns over particle containment and heating.[86] Adams maintained links to both fields: After working at the UKAEA, he returned to CERN in 1969 to oversee the construction of the Super Proton Synchrotron (SPS). By 1974, as CERN Director-General, he joined the Scientific and Technical

Committee of JET, helping to secure its approval. And when a JET director was to be appointed, Adams' deputy, Hans-Otto Wüster, was chosen.[87] Under Adams' leadership, some CERN staff also explored the application of accelerators and high-energy beam techniques to inertial confinement fusion, a dual-purpose technique that was still only partially declassified.[88] Kjell Johnsen, Norwegian member of the original fusion study group and now director of the CERN Accelerator Division, and Cornelis Zilverschoon, at CERN since 1954, believed heavy-ion beams could advance fusion science.[89] Their work caught the attention of EURATOM, and CERN was contacted to evaluate inertial confinement as a basis of considerations to collaborate on a European scale—similar as had been the case with the 1958 initiative. Not indifferently from the original group, the EURATOM Study Group for Inertial Confinement, formed in 1977, was to review activities and activate them, involving Johnsen and Zilverschoon.[90] A few years later, this became public. The authors of *La Quadrature du CERN,* a critical account of its history, argued that the CERN convention was flexible enough to accommodate basic and applied research, as illustrated through experiments with inertial confinement, including superconducting magnets and radio-frequency signals as potential reactor components.[91]

This closing episode underscores the current limited understanding of CERN's exchanges with other organisations since the 1970s, not to mention its contributions to applications beyond the World Wide Web.[92] Historiographical accounts of CERN continue to reiterate that the laboratory represents only fundamental research, as championed by its first managers. While these accounts cover the first two decades well, they fall short in addressing CERN's later evolution. They overlook CERN's growing role in applied fields, particularly through efforts to foster new avenues of collaborative research.

**Author:** *Barbara Hof, University of Lausanne, Bâtiment Anthropole, 1015 Lausanne, CH.* https://orcid.org/0000-0003-4529-544X*, e-mail:* barbara.hof@unil.ch

**Competing interests**: The author declares none.

**Acknowledgements:** I thank the Center on Technology, Society, and Humanity of the Polytechnic University of Turin (THESEUS) for supporting this work with a guest research grant. I also thank the participants of the IAEA Consultancy Meeting on the Interface Between Nuclear Science and Diplomacy, notably Matteo Barbarino, and the participants of the history seminar at Durham University, notably Joseph Martin, for their feedback. I am especially grateful to Roberto Lalli and Simone Turchetti for their helpful comments on earlier drafts and thank two anonymous reviewers for their suggestions.

[1] Svein Rosseland, 24 June 1958, CERN-ARCH-KJ-141.

[2] W. Patrick McCray, '"Globalization with Hardware": ITER's Fusion of Technology, Policy, and Politics', *History and Technology* 26, no. 4 (2010), pp. 283–312; Michel Claessens, *ITER: The Giant Fusion Reactor: Bringing a Sun to Earth*, Göttingen: Copernicus, 2019; Anna Åberg, 'The Ways and Means of ITER: Reciprocity and Compromise in Fusion Science Diplomacy', *History and Technology* 37, no. 1 (2021), pp. 106–124. Nuclear fusion has been studied as a source of stellar energy since the interwar period, but no fusion reactor has yet been developed to generate consumer energy. Advocates argue that while fusion has not yet been technologically feasible, if it can be made economically and environmentally sustainable, it could meet rising energy demands and help combat climate change, which fuels hopes of it becoming a game-changing technology.

[3] For the history of failed international institutions, see Giuliana Gemelli, 'Western Alliance and Scientific Diplomacy in the Early 1960s: The Rise and Failure of the Project to Create a European M.I.T.' In *The American Century in Europe*, edited by R. Laurence Moore and Maurizio Vaudagna, Ithaca, NY: Cornell University Press 2003, pp. 171–192; Hilary Rose, 'The Rejection of the WHO Research Centre: A Case Study of Decision-Making in International Scientific Collaboration', *Minerva* 5, no. 3 (1967), pp. 340–56.

[4] CERN, 'Convention for the Establishment of a European Organization for Nuclear Research', 1 July 1953, http://cds.cern.ch/record/480837/files/cm-p00047703.pdf

[5] Robert Lalli, 'Crafting Europe from CERN to Dubna: Physics as Diplomacy in the Foundation of the European Physical Society', *Centaurus* 63 (2021), pp. 103–131; Martin Kohlrausch and Helmuth Trischler, *Building Europe on Expertise. Innovators, Organizers, Networkers*, Basingstoke: Palgrave Macmillan, 2014, pp. 208–214.

[6] Dominique Pestre, 'Appendix to Chapter 7: Another Aspect of CERN's European Dimension: The "European Study Group on Fusion", 1958-1964,' in *History of CERN Vol. II: Building and Running the Laboratory, 1954-1965*, ed. Armin Hermann et al., Amsterdam, Oxford, New York, Tokyo: North-Holland, 1990, pp. 416–427.

[7] German A. Goncharov, 'The 50th Anniversary of the Beginning of Research in the USSR on the Potential Creation of a Nuclear Fusion Reactor', *Physics-Uspekhi* 44, no. 8 (August 2001), pp. 851–858; R.S. Pease, 'The UK Fusion Programme', *Plasma Physics and Controlled Fusion* 29 (1987), pp. 1439–1447.

[8] Matteo Barbarino, 'A Brief History of Nuclear Fusion', *Nature Physics* 16 (2020), p. 890.

⁹ The fusion process is modelled on a process that takes place in the sun. Theoretically, it is possible for various elements to fuse, but scientific considerations led to the preference of the artificial combination of deuterium (e.g., extracted from seawater) and tritium (produced from the reaction of fusion generated neutrons with naturally abundant lithium) to release energy, see John Hendry, 'The Scientific Origins of Controlled Fusion Technology', *Annals of Science* 44, no. 2 (1987), pp. 143–68; Matteo Barbarino, 'What is Nuclear Fusion?', IAEA.org, March 31, 2022, https://www.iaea.org/newscenter/news/what-is-nuclear-fusion

¹⁰ Joan Lisa Bromberg, *Fusion: Science, Politics and the Invention of a New Energy Source*, Cambridge: MIT Press, 1982, 9. The study of high-temperature plasma started earlier, but it was only with the mounting interest in fusion after 1950 that plasmas and their interactions with electromagnetic fields attracted interest, see Richard F. Post, 'Plasma Physics in the Twentieth Century'. In *Twentieth Century Physics*, edited by Laurie M. Brown, Abraham Pais, and Brian Pippard, Vol. 3, Bristol, Philadelphia, and New York: Institute of Physics Publishing and American Institute of Physics Press, 1995, pp. 1617–1690.

¹¹ Tim Flink, 'The Sensationalist Discourse of Science Diplomacy: A Critical Reflection', *The Hague Journal of Diplomacy* 15, no. 3 (2020), pp. 359–70; Charlotte Rungius and Tim Flink. 'Romancing Science for Global Solutions: On Narratives and Interpretative Schemas of Science Diplomacy', *Humanities and Social Sciences Communications* 7, no. 1 (2020), pp. 1-10. For the general argument on the relevance of competition, see Pierre-Bruno Ruffini, 'Collaboration and Competition: The Twofold Logic of Science Diplomacy', *The Hague Journal of Diplomacy* 15 (2020), pp. 371–82. For historical studies, see Jiří Janáč and Doubravka Olšáková, 'On the Road to Stockholm: A Case Study of the Failure of Cold War International Environmental Initiatives (Prague Symposium, 1971)', *Centaurus* 63, no. 1 (2021), pp. 132–149; Darina Volf, 'Evolution of the Apollo-Soyuz Test Project: The Effects of the "Third" on the Interplay Between Cooperation and Competition', *Minerva* 59, no. 3 (2021), pp. 399–418; Sam Robinson, 'Early Twentieth-Century Ocean Science Diplomacy: Competition and Cooperation among North Sea Nations', *Historical Studies in the Natural Sciences* 50, no. 4 (September 23, 2020), pp. 384–410. For its relevance in the national context, see Karin Nickelsen and Fabian Krämer, 'Introduction: Cooperation and Competition in the Sciences'. *NTM* 24, no. 2 (2016), pp. 119–23.

¹² These archival repositories have been consulted: CERN Archives at Geneva [CERN-ARCH; included are the DG-Files (Director General), DIR-ADM (Director Administration), JBA (John Bertram Adams), KJ (Kjell Johnsen), MGNH (Mervyn G.N. Hine), ISR (Intersecting Storage Rings Division)]; CERN Document Server (Council and Committee of Council, Scientific

Policy Committee, CERN Courier). UK National Archives at Kew [NA; documents from AB (Records of the United Kingdom Atomic Energy Establishment: Culham Laboratory)].

[13] R.S. Pease, 'John Adams and the Development of Nuclear Fusion Research', *Plasma Physics and Controlled Fusion* 28 (1986), p. 398.

[14] Cornelius M. Braams and Peter E. Stott, *Nuclear Fusion: Half a Century of Magnetic Confinement Fusion Research*, Bristol: Inst. of Physics Publication 2002, p. 31. For the subsequent East-West collaboration, see Climério da Silva Neto and Barbara Hof, 'Redrawing the Boundaries of Secrecy: The Anglo-Soviet Exchange in Fusion Science', In *Wissenschaft und Politik: Symposium für Christian Forstner (1975-2022)*, ed. Johannes-Geert Hagmann et al., Springer, forthcoming.

[15] Katherine Pyne, 'Art or Article? The Need for and Nature of the British Hydrogen Bomb, 1954–58'. *Contemporary Record* 9, no. 3 (1995), p. 562–585; Forna Arnold, *Britain and the H-Bomb*, Hampshire: Palgrave Macmillan, 2001.

[16] Matteo Barbarino, 'A Brief History of Nuclear Fusion', *Nature Physics* 16 (2020), p. 890.

[17] PCAST. "The U.S. Program of Fusion Energy Research and Development," July 11, 1995. https://clintonwhitehouse4.archives.gov/media/pdf/Fusion1995.pdf.

[18] J.D. Cockcroft, 'Peaceful Uses of Atomic Energy: United Nations Conference at Geneva', *Nature* 182 (1958), pp. 903–905.

[19] Conference reports by Flowers, 19 September 1958, Thompson, 19 September 1958, Bickerton, 19 September 1958, AB 73/1, NA.

[20] Bakker to Peirls, 4 March 1958, CERN-ARCH-KJ-146, as well as 'Progress Reports of the Director-General and Divisional Directors. Tenth Session of the Council', Geneva 20 June 1958, 5 June 1958.

[21] Pease, 'John Adams and the Development of Nuclear Fusion Research', p. 398.

[22] Adams, 'Study group on fusion problems', 31 July 1958, CERN-ARCH-DIR-ADM-01-DIV-PS-17.

[23] Adams' suggested P. Thonemann and R. Pease from Harwell, S. Winter, G. Vendryes, and P. Huber from Saclay, L. Biermann and A. Schlüter from Göttingen, E. Persico and B. Brunelli from Rome, C.M. Braams from Utrecht, K. Siegbahn from Uppsala, himself, Schoch and Linhart from CERN, see Adams to Bakker, 15 April 1958, CERN-ARCH-KJ-146.

[24] At the ZETA machine in Harwell, neutrons were observed when deuterons were accelerated, see Svein Rosseland, 24 June 1958, CERN-ARCH-KJ-141.

[25] H.L. Nieburg, 'EURATOM: A Study in Coalition Politics', *World Politics* 15, no. 4 (1963), pp. 597–622.

[26] Louis Armand, *Some Aspects of the European Energy Problem - Suggestions for Collective Action*, Paris: OEEC, 1955.

[27] Draft Minutes, 23 April 1958, CERN-ARCH-KJ-146.

[28] Adams to Skinner, 1 August 1957, Skinner to Cockcroft 16 August 1957, Adams to Cockcroft, 7 November 1957, Amaldi to Cockcroft, 8 November 1957, Cockcroft to Adams, 18 November 1957, all letters in AB 6/1836, NA.

[29] Armin Hermann, Laura Weiss, John Krige, Ulrike Mersits, and Dominique Pestre. *History of CERN, Vol. II: Building and Running the Laboratory, 1954-1965*, Amsterdam, Oxford, New York, Tokyo: North-Holland, 1990.

[30] 'Euratom-CERN Joint Study Group for Fusion Research, Appendix to Tenth Meeting of Council', 20 June, CERN-ARCH-KJ-146.

[31] 'Scientific Policy Committee, Draft Minutes', Geneva 23 May 1958, 1 August 1958.

[32] Scientific Policy Committee', op. cit (31).

[33] 'Press Release PR/35', 20 June 1958, as well as Guéron to Bakker, 12 June 1958, Bakker to Guéron, 24 June 1958, CERN-ARCH-KJ-146.

[34] 'Minutes, Tenth Session of the Council, 20-27 June', approved 9 October 1958.

[35] Mervyn O'Driscoll, 'Missing the Nuclear Boat? British Policy and French Military Nuclear Ambitions During the Euratom Foundation Negotiations, 1955–56', *Diplomacy and Statecraft* 9, no. 1 (1998): 135–62; John Krige, 'The Peaceful Atom as Political Weapon: Euratom and American Foreign Policy in the Late 1950s', *Historical Studies in the Natural Sciences* 38, no. 1 (2008): 5–44; Simone Turchetti, 'A Most Active Customer: How the U.S. Administration Helped the Italian Atomic Energy Project to "De-Develop."' *Historical Studies in the Natural Sciences* 44, no. 5 (2014): 470–502.

[36] Swiss Delegation, 'Draft Statement by Member States of CERN which do not belong to Euratom. Tenth Session of the Council', 20 June 1958. Details in 'Fusion research: Proposed setting up of informal study group. Tenth Session of the Council', 25/27 June 1958.

[37] Huet to Directeur Général, 18 June 1958, CERN-ARCH-KJ-14.

[38] D.W. Fry to Patrick Blackett, 24 June 1958, AB 6/1982, NA.

[39] 'Enclosure', M. Verry to Bakker, 26 June 1958, CERN-ARCH-KJ-146.

[40] 'CERN and fusion research, comments by Advisor to the UK delegation', H.L. Verry, 26 March 1958, AB 6/1982, NA.

[41] H.W. Melville to Cockcroft, 31 March 1958, AB 6/1982, NA.

[42] D.E.H. Peirson to J. D. Cockcroft, 28 November 1957, J.D. Cockcroft to D.E.H. Peirson, 29 November 1957, AB 6/1982, NA.

[43] 'Plasma physics in CERN' John Cockcroft, 12 July 1958, AB 6/1982, NA.

[44] Cockcroft to Peirson, 26 June 1958, AB 6/1982, NA.

[45] 'Draft Resolution proposed by the Danish Delegation. Tenth Session of the Council', 27 June 1958.

[46] 'Draft Resolution proposed by the British Delegation. Tenth Session of the Council', 27 June 1958; 'Amendments to Draft Resolution proposed by the U.K. Delegation. Tenth Session of the Council', 27 June 1958; 'Euratom-CERN joint study group for fusion research, resolution adopted by the Council. Tenth Session of the Council', 10 July 1958.

[47] 'Confidential note', H.L. Verry to Secretary, 1 July 1958, AB 6/1982, NA.

[48] John Krige, 'Britain and the European Laboratory Project Mid-1952-December 1953', in *History of CERN, Vol I: Launching the European Organization for Nuclear Research*, edited by Armin Hermann, John Krige, Ulrike Mersits, and Dominique Pestre, (Amsterdam, Oxford, New York, Tokyo: North-Holland, 1987), p. 497.

[49] Bakker to Guéron, 3 July 1959, as well as Guéron to Bakker, 9 July 1958, Adams to Guéron, 29 July 1958, CERN-ARCH-KJ-146.

[50] D.W. Fry to Director, Draft 5 December 1958, AB 6/1982, NA. There is no documented participation of delegates from CERN member states Greece and Yugoslavia. Austria joined CERN in 1959 and sent delegates, see list name list in 'Steering Committee for Nuclear Energy, Work of the CERN study group of fusion', ENEA, 4 June 1959, AB 6/1982.

[51] Kjell Johnsen to J.B. Adams, 21 August 1958, CERN-ARCH-KJ-141.

[52] Adams to P. Huet, 29 July 1958, CERN-ARCH-KJ-146; Note by the Secretary of ENEA, 4 June 1959, Annex to Note by the Secretary of ENEA, 4 June 1959, AB 6/1982, NA.

[53] 'First meeting of CERN study group on fusion', 13 October 1958, CERN-ARCH-MGNH-077.

[54] Bakker to McKinney, 25 March 1960, CERN-ARCH-DG-FILES-170; 'Manpower and expenditures in world programs', 1963, p. 6, AB 82/8, NA.

[55] 'Meeting summary CERN fusion study group by R. Bickerton', 11-12 December 1958, AB 6/1982, NA.

[56] 'European Fusion Research, Report of the CERN Study Group on Fusion Problems', 14 April 1959, CERN-ARCH-DIR-ADM-01-DIV-PS-17. Also available at https://cds.cern.ch/record/17803/files/CM-P00076290-e.pdf

[57] Johnsen to Adams, 20 February 1959, CERN-ARCH-KJ-143.

[58] Leading board, 15 April 1959, CERN-ARCH-DIR-ADM-01-DIV-PS-17. This provided the basis of Bakker's position: 'Comments by the Director-General, Thirteenth Session of the Council', 26 May 1959.

[59] In the following year it was clear that a new Brookhaven accelerator provided higher energies than those at CERN. Consequently, the Council decided to design new accelerators, which later resulted in the creation of the Intersecting Storage Rings (ISR) and the Super Proton Synchrotron (SPS), see 'CERN Councils', *CERN Courier* July 1960: 8.

[60] 'Draft Minutes, Thirteen Session of the Council', 26 May 1959.

[61] 'The Future of the CERN Study Group on Fusion Problems', Adams, 14 October 1959, CERN-ARCH-DG-FILES-169; Details in 'European Society for Controlled Thermonuclear Fusion Research, Statutes', Draft on 2 March 1959, CERN-ARCH-DG-FILES-170.

[62] 'CERN Study Group on Fusion Problems, Fourteenth Session of the Council on 2 December 1959', 29 October 1959.

[63] Bakker to Dakin, 13 October 1959, CERN-ARCH-DG-FILES-170.

[64] 'Sixth meeting of the European Study Group on Fusion', 11 October 1960, CERN-ARCH-KJ-146.

[65] 'European Study Group on Fusion, sponsored by CERN, Geneva, Attached Report, Twenty-First Session of the Council', Adams, 5 June 1962, CERN-ARCH-DIR-ADM-01-DIV-PS-17.

[66] E.N. Shaw, *Europe's Nuclear Power Experiment History of the OECD Dragon Project,* Oxford, New York, Toronto, Sydney, Paris, Frankfurt: Pergamon Press, 1982.

[67] 'CERN Study Group on Fusion, Second Meeting', CERN/FSG/6, 16 December 1958, CERN-ARCH-KJ-142, here pp. 4-6.

[68] Barbara Curli, "The Origins of Euratom's Research on Controlled Thermonuclear Fusion: Cold War Politics and European Integration, 1958–1968," *Contemporary European History*, 2022, pp. 1–19.

[69] J. Adams to J. Cockcroft, 1 October 1959, J.F. Jackson to J.C. Walker, 31 July 1959, AB 6/1982, NA.

[70] Alex Wellerstein, *Restricted Data. The History of Nuclear Secrecy in the United States*, Chicago: Chicago University Press, 2021; Kristan Stoddart, 'British Nuclear Strategy During the Cold War'. In *The British Way in in Cold Warfare. Intelligence, Diplomacy, and the Bomb, 1945-1975*, edited by Matthew Grant, London, New York: Continuum, 2009, pp. 19-20.

[71] Minutes of the meeting of the C.T.R. Advisory Committee on 4 November 1959, AB 73/1, as well as Culham staffing policy, 19 January 1960, AB 73/4, and first meeting of Scientific Planning Committee held 4 October 1961, AB 73/16, all in NA.

[72] '"Potted" history of C.T.R. Project', AB 77/1, NA.

[73] John Adams to Don W. Fry, 25 November 1958 and C.T.R. Advisory Committee, 15 November 1962, AB 6/1982, NA.

[74] 'Draft Reorganisation of CTR work at Harwell', W.G. Penney, 22 February 1960, as well as Adams to W.G. Penney, 10 February 1960, AB 77/2 and 'Follow up of work done during October visit of JBA', AB 77/8, all in NA.

[75] Willson to Director, 22 January 1960, AB 73/4, NA; 'CERN Councils', *CERN Courier* July 1960: 8.

[76] 'Minutes of the Eighteenth Session of the Council on 8 and 9 December 1960', CERN-ARCH-DIR-ADM-01-DIV-PS-17.

[77] S. Dakin to J. Adams, 9 November 1961, AB 77/22, NA; and J. Adams to S. Dakin, 13 November 1961, CERN-ARCH-DG-FILES-170.

[78] Matteo Barbarino, 'Past, Present and Future of Fusion Science Diplomacy,' *Communications Physics*, 2021, p. 2.

[79] 'Draft Minutes Twentieth Session of the Council', 11 June 1962, CERN-ARCH-DG-FILES-170.

[80] 'Future meetings of the European Study Group on Fusion', Adams, undated, CERN-ARCH-JBA-187.

[81] Adams to Speeckaert, 8 January 1964, AB 77/22, NA.

[82] Adams to Pease, 12 July 1968, Pease to Adams, 15 July 1968, AB 77/43, NA.

[83] D.E.H. Peirson to R. Arculus, 17 April 1970, AB 77/43, NA

[84] R.S. Pease and D.R. Willson, 'International Collaboration in Research on Controlled Thermonuclear Fusion', *Contemporary Physics* 15, no. 2 (1974), pp. 179–92.

[85] E.N. Shaw, *Europe's Experiment in Fusion: The JET Joint Undertaking*, Amsterdam: North-Holland, 1990. JET achieved its first plasma in 1983 and is now decommissioned again. It provided important insights for the later ITER.

[86] European Study Group on Fusion, op. cit (64).

[87] 'Joining the JET Set', *CERN Bulletin*, no. 4 (1978), p. 1.

[88] Relevant for the partial declassification was a conference in Montreal in 1972, see Johannes-Geert Hagmann, Licht und Laserphysik. In J. Renn, C. Reinhardt, J. Kocka, F. Schmaltz, B. Kolboske, J. Balcar, et al. (Eds.), *Die Max-Planck-Gesellschaft: Wissenschafts- und Zeitgeschichte 1945–2005*. Göttingen: Vandenhoeck & Ruprecht, 2024, p. 326.

[89] 'Serpukhov Accelerator Conference', *CERN Courier* 17, no. 7–8 (1977), pp. 228–33.

[90] R. Balescu to Kjell Johnsen, 21 December 1976, and 'Minutes of the First Meeting of the Euratom Study Group for Inertial Confinement', 1 April 1977, CERN-ARCH-ISR-03-1-011; 'Ion accelerators for fusion, ISR seminar', Kjell Johnsen, 20 June 1977, CERN-ARCH-KJ-082; 'New approaches to fusion research: heavy ions', Kjell Johnsen, 30 September 1977, CERN-ARCH-KJ-088.

[91] Jacques Grinevald et al., *La Quadrature Du CERN: Essai indisciplinaire publié à l'occasion du 30e Anniversaire du CERN*, Lausanne: Edition d'en bas, 1984, pp. 19-21.

[92] Bebo White, 'The World Wide Web and High-Energy Physics', *Physics Today* 51, no. 11 (1998): 30-36.